# NFV Orchestrator Placement for Geo-Distributed Systems

Mohammad Abu-Lebdeh, Diala Naboulsi, Roch Glitho
*Concordia University*
Montreal, Canada
{m_abuleb, d_nabou, glitho}@encs.concordia.ca

Constant Wette Tchouati
*Ericsson*
Montreal, Canada
constant.wette.tchouati@ericsson.com

*Abstract*—The European Telecommunications Standards Institute (ETSI) developed Network Functions Virtualization (NFV) Management and Orchestration (MANO) framework. Within that framework, NFV orchestrator (NFVO) and Virtualized Network Function (VNF) Manager (VNFM) functional blocks are responsible for managing the lifecycle of network services and their associated VNFs. However, they face significant scalability and performance challenges in large-scale and geo-distributed NFV systems. Their number and location have major implications for the number of VNFs that can be accommodated and also for the overall system performance. NFVO and VNFM placement is therefore a key challenge due to its potential impact on the system scalability and performance. In this paper, we address the placement of NFVO and VNFM in large-scale and geo-distributed NFV infrastructure. We provide an integer linear programming formulation of the problem and propose a two-step placement algorithm to solve it. We also conduct a set of experiments to evaluate the proposed algorithm.

*Index Terms*—Distributed systems, network functions virtualization, management and orchestration, NFV orchestrator, VNF manager, scalability, placement.

## I. INTRODUCTION

Network Functions Virtualization (NFV) induces profound changes in the architecture of fixed and mobile networks. It decouples the network function software from underlying hardware allowing the software to run on commodity hardware. By that, NFV provides the necessary flexibility to enable agile, cost-effective, and on-demand service delivery model combined with automated management.

The European Telecommunications Standards Institute (ETSI) defined the NFV architectural framework [1] as depicted in Fig. 1. It details the key functional blocks in NFV system: Virtualized Network Function (VNF), NFV Infrastructure (NFVI) and NFV Management and Orchestration (MANO) framework [2]. The latter comprises of the Virtualized Infrastructure Manager (VIM), VNF Manager (VNFM) and NFV Orchestrator (NFVO). The communication between various functional blocks in NFV architectural framework is enabled through a set of well-defined reference points.

The VNF is the software implementation of a network function. It runs on NFV Infrastructure (NFVI) that encompasses the necessary software and hardware components. NFVI can span geographically distributed locations, called NFVI Point of Presences (NFVI-PoPs). NFVI resources (e.g. compute, storage and network) can be managed and controlled by one or more VIMs. Moreover, the VNFM is responsible for the lifecycle management (e.g. instantiation, scaling, healing and monitoring) of one or more VNF instances. The NFVO is in charge of the orchestration of NFVI resources across multiple VIMs and lifecycle management of network services. The NFVO and VNFM work jointly to ensure that the network services and their corresponding VNFs meet the service quality requirements (e.g. reliability). Thus, the performance of NFVO and VNFM is crucial to NFV system.

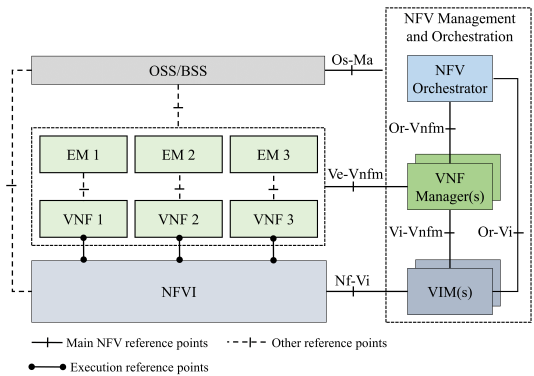

Fig. 1. ETSI NFV architectural framework.

In many emerging NFV scenarios, the network services will span large geographical area (e.g. country or continent) and the number of VNF instances will grow tremendously. For instance, the forthcoming 5G cellular system, for which NFV is considered an essential enabling technology, is likely to run on a highly distributed NFVI to achieve the requirement of $1\,\mathrm{ms}$ round-trip latency. 5G also requires massively scalable NFV architecture [3] to deliver the required massive capacity and connectivity. In these scenarios, the performance of NFVO and VNFM is a critical challenge due to its impact on the NFV system. As shown in Fig. 1, one or more VNFMs can exist in an administrative domain (e.g. service provider domain), whereas there is a single NFVO. However, considering a large-scale and distributed NFV system, relying on a single NFVO in the system will hinder the scalability of the orchestration process [4]. The number of VNFMs should also be adequate for managing the VNF instances. Besides, the location of NFVO and VNFM play a vital role in determining the communication delay over their reference points. Therefore, the number of NFVOs and VNFMs along with their placement should be planned to ensure the scalability and performance of NFV management and orchestration.

Moreover, hierarchical service orchestration has recently been employed in NFV to address various challenges such as the orchestration across multi-technology domains. We believe that the same concept can be used to address the NFV MANO scalability and performance challenges for large-scale and distributed NFV systems. In this paper, we consider an architecture of two layers of network service orchestration that enables the existence of multiple NFVOs in an operator environment. Then, in this context, we tackle the challenge of identifying the optimal number and location of NFVOs and VNFMs needed to manage a given set of VNF instances distributed over a set of NFVI-PoPs. We propose an Integer Linear Programming (ILP) formulation of the problem. The formulation accounts for the capacity (e.g. NFVO and VNFM) and delays between various functional blocks to satisfy the scalability and performance requirements of the system. In addition, we propose a two-step placement algorithm and evaluate it.

The rest of paper is organized as follows. We start by discussing related work in section II. Section III is dedicated to system architecture and problem description. We present the resolution approach in section IV, followed by the performance evaluation in section V. Finally, we draw our conclusions in section VI.

## II. RELATED WORK

### A. Hierarchical NFV Orchestration

Several studies employ hierarchical orchestration of network service in NFV systems. However, to the best our knowledge, they address various architectural challenges. In contrary, this work tackles algorithmic aspects and aim to optimize the placement of NFVO and VNFM. For example, ETSI report [5] presents a use case of using two layers of NFVOs to offer a network service across two administrative domains. Garay et al. [4] propose a novel service graph model that enables the hierarchical orchestration of a network service. The European 5G Exchange (5GEx) project [6] proposes a platform to enable service orchestration over multiple domains for the same or different administrations in the context of 5G.

### B. Resource Allocation in NFV

The VNF placement problem has received substantial attention in the literature [7]. Formally, the problem is defined as selecting the locations for a chain of VNF instances. A chain of VNF instances is a set of VNF instances linked together to provide a network service. There are multiple objectives to consider with VNF placement. For example, the authors in [8] aim at minimizing the operational cost of the system. Mechtri et al. [9] consider the efficiency of resource utilization. Hirwe and Kataoka [10] focus on minimizing the length of paths traversed by flows. As per definition, the VNF placement problem pays no attention to NFV MANO functional blocks such as NFVO and accounts only for the VNF functional block.

Furthermore, to the best of our knowledge, our previous work [11] is the only one that investigates the resource allocation of NFV MANO functional blocks. In that study, we propose an online algorithm to optimize the VNFM placement over a distributed NFVI. The algorithm aims at minimizing the operational cost while ensuring the performance requirements (e.g. VNFM capacity). However, the proposed algorithm considers only a single NFVO placed at a given location.

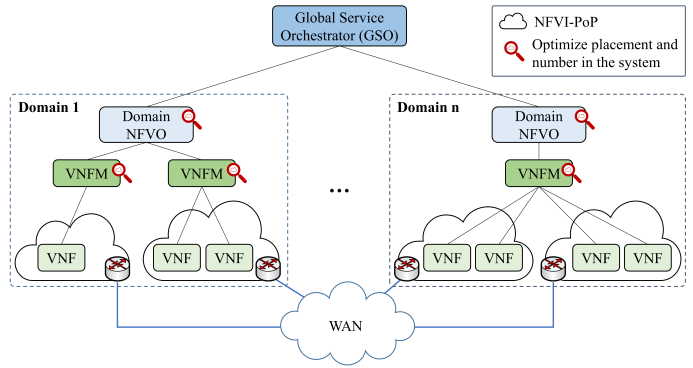

Fig. 2. High-level system architecture.

## III. SYSTEM ARCHITECTURE AND PROBLEM DESCRIPTION

As shown in Fig. 2, we consider a distributed NFVI consisting of multiple NFVI-PoPs located in different regions and interconnected through a WAN network. The architecture supports two-layer hierarchical service orchestration. The bottom layer encompasses a set of NFVOs; each performs the same functionality as defined by ETSI MANO framework. Their number will depend on the capacity and delay requirements of the system. At the top layer, there is a Global Service Orchestrator (GSO) responsible for the end-to-end service orchestration across multiple NFVOs.

It is important to note that the architecture is fully aligned with the ETSI MANO framework. The NFVI is decomposed into a set of domains such that their number is equivalent to the number of NFVOs. Each domain consists of one or more NFVI-PoPs. Within a domain, there are a single NFVO and one or more VNFMs that perform the standard functions defined by ETSI MANO framework.

The architecture enables the system to scale the number of NFVOs and VNFMs and meet the scalability and performance requirements of different scenarios. However, there is still a challenge of finding their number and placement. We formally define the problem as follows: given a set of NFVI-PoPs, a set of VNF instances along with their location, and fixed location of the GSO, then our goal is to find (1) the optimal number and placement of NFVOs needed in the system, (2) the NFVI-PoPs assigned to each NFVO, i.e. NFVI-PoPs in each domain, (3) the number and placement of VNFMs in each domain, and (4) the VNF instances assigned to each VNFM. Our objective is to minimize the number of NFVOs and VNFMs while fulfilling the capacity (e.g. NFVO and VNFM) and delay constraints.

Furthermore, for the sake of simplicity, we make the following assumptions: (1) a VNF instance and its Element Manager (EM) are deployed at the same NFVI-PoP, (2) a VIM manages resources within one NFVI-PoP, and it is placed at that NFVI-

PoP, and (3) all functional blocks communicate over the same network links, i.e. no dedicated links.

*A. System Model*

Consider the NFVI modeled as a graph $G = (P, E)$ where $P$ is the set of NFVI-PoP nodes and $E$ is the set of edges linking them, such that $E = \{(p, q) \mid p \in P, q \in P, p \neq q\}$. We use $\delta_{p,q}$ to represent the network delay of an edge $(p, q) \in E$. Let $V$ represent the set of VNF instances in the system. The location of a VNF instance $v \in V$ is defined by $l_{v,p} \in \{0, 1\}$ such that $l_{v,p}$ equals to 1 only when $v$ is placed at $p \in P$. We define $M$ to represent the set of VNFMs that can be used to manage the VNF instances. We also use $\varphi$ to denote the capacity of a VNFM. It represents the maximum number of VNF instances that can be managed by a VNFM.

We consider that a NFVO has capacity defined in terms of the maximum number of VNF instances in its domain. We employ $\Phi$ to refer to this capacity. Moreover, we assume that the GSO is deployed at a given NFVI-PoP. We define $w_p \in \{0, 1\}$ to indicate the GSO location, such that $w_p$ is equal to 1 only if the GSO is placed at $p \in P$.

We consider that there is an upper bound on the acceptable network delay between various functional blocks to ensure predictable system performance. We use $\psi$ and $\Psi$ to denote the maximum acceptable delay between a NFVO on the one hand, the GSO and a VIM on the other hand. Moreover, the same VNFM can manage different VNF types (e.g. firewall) which can impose different requirements on the network delay over the VNFM reference points. Thus, we define the upper bound on network delay between a VNFM and other functional blocks per VNF instance. We use $\Omega_v$ to indicate the maximum acceptable delay between the NFVO and the VNFM assigned to VNF instance $v$. We also employ $\omega_v$ to denote the upper bound on the delay between the VNF instance $v$ and its designated VNFM. Due to our assumptions (2) and (3), $\omega_v$ also represents the maximum acceptable delay between the VNFM and the VIM of NFVI-PoP where $v$ is located.

*B. Problem Formulation*

We mathematically formulate the problem as an ILP model as presented below.

**Decision Variables:**
$h_p \in \{0, 1\}$ : (1) indicates that a NFVO is placed at $p \in P$, (0) otherwise.
$r_{q,p} \in \{0, 1\}$ : (1) specifies that $q \in P$ is assigned to the NFVO which is placed at $p \in P$, (0) otherwise.
$x_{m,p} \in \{0, 1\}$ : (1) designates that $m \in M$ is placed at $p \in P$, (0) otherwise.
$y_{v,m,p} \in \{0, 1\}$ : (1) indicates that $v \in V$ is assigned to $m \in M$ which is placed at $p \in P$, (0) otherwise.

**Mathematical Model:**

$$\text{Minimize} \quad \sum_{p \in P} h_p + \sum_{m \in M} \sum_{p \in P} x_{m,p} \quad (1)$$

$$\text{Subject to:} \quad \sum_{p \in P} r_{q,p} = 1, \quad \forall q \in P \quad (2)$$

$$r_{q,p} \leq h_p, \quad \forall q, p \in P \quad (3)$$

$$r_{p,p} = h_p, \quad \forall p \in P \quad (4)$$

$$\sum_{p \in P} x_{m,p} \leq 1, \quad \forall m \in M \quad (5)$$

$$\sum_{m \in M} \sum_{p \in P} y_{v,m,p} = 1, \quad \forall v \in V \quad (6)$$

$$y_{v,m,p} \leq x_{m,p}, \quad \forall v \in V, m \in M, p \in P \quad (7)$$

$$l_{v,q} \, y_{v,m,\acute{p}} \, r_{\acute{p},p} \leq r_{q,p}, \quad \forall v \in V, m \in M, q, \acute{p}, p \in P \quad (8)$$

$$\sum_{v \in V} \sum_{m \in M} \sum_{q \in P} y_{v,m,q} \, r_{q,p} \leq \Phi \, h_p, \quad \forall p \in P \quad (9)$$

$$\sum_{v \in V} y_{v,m,p} \leq \varphi \, x_{m,p}, \quad \forall m \in M, p \in P \quad (10)$$

$$x_{m,p} \leq \sum_{v \in V} y_{v,m,p}, \quad \forall m \in M, p \in P \quad (11)$$

$$w_p \, h_q \, \delta_{p,q} \leq \psi, \quad \forall (p, q) \in E \quad (12)$$

$$r_{q,p} \, \delta_{p,q} \leq \Psi, \quad \forall (p, q) \in E \quad (13)$$

$$l_{v,p} \, y_{v,m,q} \, \delta_{p,q} \leq \omega_v, \quad \forall v \in V, m \in M, (p, q) \in E \quad (14)$$

$$y_{v,m,q} \, r_{q,p} \, \delta_{p,q} \leq \Omega_v, \quad \forall v \in V, m \in M, (p, q) \in E \quad (15)$$

The objective function (Eq.1) seeks at minimizing the number of NFVOs and VNFMs as their number is a measure of the operational cost of the NFV management and orchestration. Constraint (2) stipulates that one NFVO is responsible for the resource orchestration of a NFVI-PoP, i.e. a NFVI-PoP belongs exactly to one domain. Constraint (3) ensures that the NFVI-PoP $q$ can be assigned to the NFVO at NFVI-PoP $p$ when there exists an active NFVO at $p$. Constraint (4) indicates that a NFVO should be placed within its domain boundaries. Constraint (5) ensures that a VNFM can be placed only at one NFVI-PoP. Constraint (6) indicates that each VNF instance should be assigned to one VNFM. Constraint (7) stipulates that a VNF instance can be assigned to VNFM $m$ placed at NFVI-PoP $p$ only when $m$ exists at $p$. Constraint (8) indicates that a VNF instance and its designated VNFM should exist in the same domain. We enforce that capacity constraints of NFVO and VNFM by (9) and (10). Constraint (11) ensures that a VNFM is active only when it manages at least one VNF instance. Constraints (12)-(15) enforce the delay limits in the system.

Note that the constraints (8), (9) and (15) are non-linear constraints and can be linearized by replacing them with linear constraints (16)-(21) as follows.

$$l_{v,q} \, z_{v,m,\acute{p},p} \leq r_{q,p}, \quad \forall v \in V, q, \acute{p}, p \in P \quad (16)$$

$$\sum_{v \in V} \sum_{m \in M} \sum_{q \in P} z_{v,m,q,p} \leq \Phi \, h_p, \quad \forall p \in P \quad (17)$$

$$z_{v,m,q,p} \, \delta_{p,q} \leq \Omega_v, \quad \forall v \in V, m \in M, (p, q) \in E \quad (18)$$

$$z_{v,m,q,p} \leq y_{v,m,q}, \quad \forall v \in V, m \in M, (p, q) \in E \quad (19)$$

$$z_{v,m,q,p} \leq r_{q,p}, \quad \forall v \in V, m \in M, (p, q) \in E \quad (20)$$

$$z_{v,m,q,p} \geq y_{v,m,q} + r_{q,p} - 1, \quad \forall v \in V, m \in M, (p, q) \in E \quad (21)$$

**Algorithm 1:** Two-Step Placement Algorithm

```
  /* Step One                                            */
1 h_p, r_{q,p} ⟵ call NFVO_Placement()
  /* Step Two                                            */
2 foreach p ∈ P do
3    if h_p = 1 then
4        compute VNFM placement inputs for this domain
5        S ⟵ call VNFM_Placement()
6        x_{m,p}, y_{v,m,p} ⟵ extract decision variables from solution S
7    end
8 end
9 return (h_p, r_{q,p}, x_{m,p}, y_{v,m,p})
```

## IV. RESOLUTION APPROACH

We propose Two-Step Placement (TSP) algorithm to solve the overall problem, as illustrated in Algorithm 1. It begins by first decomposing the NFVI into one or more domains and placing a single NFVO in each domain. This step is performed through a tabu search based algorithm which will be discussed later in section IV-A. However, in general, the algorithm aims to minimize the number of NFVOs in the system. It gives the solution of decision variables $h_p$ and $r_{q,p}$ that satisfies the model constraints (2)–(4), (12) and (13). Besides, this step disregards the placement of VNFMs themselves. However, we assure that the solution would give the possibility for future VNFM placement to satisfy the VNFM delay constraints, i.e. constraints (14) and (15). To do so, we impose additional constraints on the solution to ensure that $\forall v \in V, \exists \acute{p} \in P$ such that:

$$l_{v,q}\, r_{q,p}\, r_{\acute{p},p}\, \delta_{q,\acute{p}} \leq \omega_v, \quad \forall q, p \in P \quad (22)$$

$$l_{v,q}\, r_{q,p}\, r_{\acute{p},p}\, \delta_{\acute{p},p} \leq \Omega_v, \quad \forall q, p \in P \quad (23)$$

The constraints (22) and (23) guarantee that for every VNF instance $v$, there exists a NFVI-PoP $\acute{p}$ in the same domain where a VNFM can be placed to manage $v$ while fulfilling the delay constraints. After that in the second step, for each domain, we place the needed VNFMs and map the VNF instances onto the VNFMs. We do that by utilizing the VNFM placement algorithm presented in our earlier work [11]. This step provides the solution of decision variables $x_{m,p}$ and $y_{v,m,p}$. The obtained solution satisfies the model constraints (5)–(11), (14) and (15).

### A. NFVO Placement Algorithm

We propose a tabu search based algorithm to place the NFVOs and define the boundaries of their domains. Tabu search [12] is a widely adopted meta-heuristic that guides a local search procedure to explore the solution space beyond local optimality. It starts the search process from an initial solution and iteratively explores the neighbor solutions. In each iteration, it uses movements to produce a set of neighbor solutions and employs an objective function to evaluate them. The search continues till the stop criteria are met. In what follows, we present the key components of the proposed tabu search algorithm:

*1) Initial solution:* The algorithm starts with a simple initial solution where a NFVO is placed at each NFVI-PoP in the system. The resulting solution may be infeasible, violating the delay constraint between GSO and NFVO. However, it provides a good enough starting solution that tabu search can improve gradually.

*2) Neighborhood structure:* We define two movements to transit from the current solution to a neighbor solution. The first one is to select a NFVO randomly and then invert its state, i.e. change from active to inactive and vice versa. The second movement is to draw a NFVI-PoP randomly and reassign it to another NFVO chosen at random.

*3) Acceptance criteria:* We relax the constraints (2)–(4), (12), (13), (22) and (23) to allow tabu search to explore the infeasible boundary. However, we assign a penalty for each solution to lead the algorithm to satisfy those constraints through the search process. We use two objectives in scoring a solution: solution penalty and number of NFVOs in the system. The algorithm aims first to minimize the solution penalty, then the number of NFVOs. We also employ a simple oscillation strategy in the solution evaluation. In each iteration, if there is a neighbor solution that has a better score than the best-found solution, then we choose it. Otherwise, we select a solution that minimizes the number of NFVOs, although it may not have the lowest penalty. Our goal is to drive the search to explore the infeasible solutions and thus induce diversification.

*4) Stop criteria:* The algorithm stops after $(4 \times |P|)$ consecutive iterations without an improvement in the solution. The formula allows the number of iterations to grow with respect to the number of NFVI-PoPs. The multiplier 4 is adjusted experimentally.

## V. EVALUATION

In this section, we compare the performance of TSP algorithm with the optimal solution obtained by solving the ILP model with CPLEX. In the following, we first describe the simulation setup, followed by comparison results. The simulation parameters are listed for convenience in Table I.

### A. Simulation Setup

We simulate two different NFVIs with 8 and 16 NFVI-PoPs. Each NFVI-PoP represents an AT&T data center located in North America [13]. The inter NFVI-POP delays are the round-trip delay of ping packets between each pair of NFVI-PoPs and obtained from public ping statistics [14]. The GSO is placed at a NFVI-PoP with a central location in the NFVI structure. We also consider that the capacity of a NFVO and a VNFM are 20 and 10 VNF instances, respectively. In our experiments, the VNF instances are placed uniformly at random over NFVI-PoPs, and their number varies from 10 to 60. Further, we assume that communication between the GSO and NFVO tolerates higher network delay compared to the communication between the functional blocks inside a domain. We further consider that the communication between the VNF instance and its designated VNFM is more sensitive to delay compared to the communication between other functional blocks inside a domain. Table I presents the maximum acceptable delay between various functional used in our simulation.

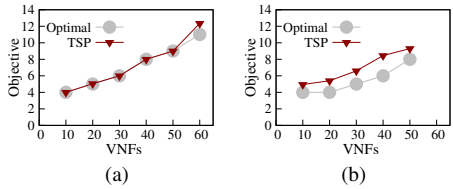

Fig. 3. Objective function value for (a) 8 NFVI-PoPs, and (b) 16 NFVI-PoPs.

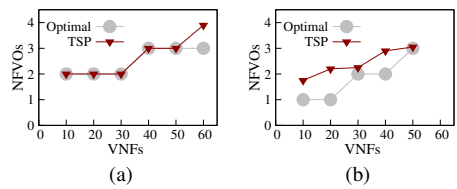

Fig. 4. Number of NFVOs for (a) 8 NFVI-PoPs, and (b) 16 NFVI-PoPs.

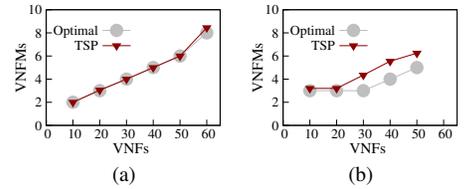

Fig. 5. Number of VNFMs for (a) 8 NFVI-PoPs, and (b) 16 NFVI-PoPs.

TABLE I
SIMULATION PARAMETERS

| Parameter | Value |
| --- | --- |
| Number of NFVI-PoPs ($|P|$) | 8, 16 |
| GSO location ($w_p$) | Central NFVI-PoP |
| NFVO capacity ($\Phi$) | 20 |
| VNFM capacity ($\varphi$) | 10 |
| Number of VNF instances ($|V|$) | 10–60 |
| Acceptable delay between GSO and NFVO ($\psi$) | 80 ms |
| Acceptable delay between NFVO and VIM ($\Psi$) | 60 ms |
| Acceptable delay between NFVO and VNFM managing VNF instances $v$ ($\omega_v$) | 45 ms |
| Acceptable delay between VNFM and VNF instance $v$ ($\Omega_v$) | 30 ms |

### B. Results

Fig. 3 portrays the objective function value, i.e. the total number of NFVOs and VNFMs, of the optimal and TSP solutions for 8 and 16 NFVI-PoPs. The TSP results are the average of 20 runs of each experiment. In Fig. 3(a), we observe that TSP provides solutions that are very close to the optimal solution and attains the optimality in most cases. However, Fig. 3(b) shows that TSP gives solutions that are within 1.4 times of the optimal solutions. Further, the results indicate that the number of NFVOs and VNFMs increases gradually with the growth of VNF instances in the system.

For a better interpretation of the results, we provide the detailed number of NFVOs and VNFMs in Fig. 4 and Fig. 5 respectively. We can easily see that the number of VNFMs grows at a higher rate compared to the number of NFVOs. The main reason is that a NFVO can accommodate more VNF instances than a VNFM as it has higher capacity. In general, the results point out that the NFV MANO capacity is adjusted to accommodate the existing VNF instances.

Moreover, although a single NFVO has adequate capacity to accommodate 10 VNF instances in our experiments, interestingly Fig. 4(a) reports that two NFVOs are needed when the NFVI consists of 8 NFVI-PoPs, whereas Fig. 4(b) tells that a single NFVO is sufficient for 10 VNF instances over 16 NFVI-PoPs. This difference in the results is attributed to the NFVI-PoPs distribution and number. The system with 8 NFVI-PoPs is constrained to a small number of NFVI-PoPs which imposes the need of additional NFVO to satisfy the delay constraints in the system, whereas the NFVI of 16 NFVI-PoPs provides the system with a higher degree of flexibility and allows fulfilling the delay constraints using one NFVO. Considering Fig. 5, we notice that the system needs 2 and 3 VNFMs to manage 10 VNF instances over 8 and 16 NFVI-PoPs respectively. We can thus conclude that number of NFVI-PoPs and their locations impact the number of NFVOs and VNFMs needed in the system.

## VI. CONCLUSION

In this paper, we motivated the adoption of hierarchical service orchestration architecture to overcome the scalability and performance challenges of NFV management and orchestration. The architecture enables the system to scale out the number of NFVOs and VNFMs and meet various scalability and performance requirements. In this context, we addressed the problem of finding the optimal number of NFVOs and VNFMs along with their placement. We formulated the problem as ILP and proposed two-step placement algorithm. We used two NFVI topologies to evaluate our algorithm. The numerical result indicated that the number of NFVI-PoPs has an impact on the required number of NFVOs and VNFMs in the system.

## REFERENCES

[1] ETSI, "Network Functions Virtualization (NFV); Architectural Framework," Dec. 2014.
[2] ETSI, "Network Functions Virtualisation (NFV); Terminology for Main Concepts in NFV," Dec. 2014.
[3] "Network Operator Perspectives on NFV priorities for 5g," Tech. Rep.
[4] J. Garay, J. Matias, J. Unzilla, and E. Jacob, "Service description in the NFV revolution: Trends, challenges and a way forward," *IEEE Commun. Mag.*, vol. 54, no. 3, pp. 68–74, Mar. 2016.
[5] ETSI, "Network Functions Virtualisation (NFV); Management and Orchestration; Report on Architectural Options," Jul. 2016.
[6] C. J. Bernardos, B. P. Ger, M. Di Girolamo, A. Kern, B. Martini, and I. Vaishnavi, "5gex: realising a Europe-wide multi-domain framework for software-defined infrastructures," *Trans. Emerging Tel. Tech.*, vol. 27, no. 9, pp. 1271–1280, Sep. 2016.
[7] J. G. Herrera and J. F. Botero, "Resource allocation in nfv: A comprehensive survey," *IEEE Trans. Netw. Service Manag.*, vol. 13, no. 3, pp. 518–532, 2016.
[8] C. Pham, N. H. Tran, S. Ren, W. Saad, and C. S. Hong, "Traffic-aware and energy-efficient vnf placement for service chaining: Joint sampling and matching approach," *IEEE Trans. Mobile Comput.*, 2017.
[9] M. Mechtri, C. Ghribi, and D. Zeghlache, "A Scalable Algorithm for the Placement of Service Function Chains," *IEEE Trans. Netw. Service Manag.*, vol. 13, no. 3, pp. 533–546, Sep. 2016.
[10] A. Hirwe and K. Kataoka, "Lightchain: A lightweight optimisation of vnf placement for service chaining in nfv," in *NetSoft Conference and Workshops (NetSoft), 2016 IEEE*. IEEE, 2016, pp. 33–37.
[11] M. Abu-Lebdeh, D. Naboulsi, R. Glitho, and C. W. Tchouati, "On the Placement of VNF Managers in Large-Scale and Distributed NFV Systems," *IEEE Trans. Netw. Service Manag.*, vol. PP, no. 99, pp. 1–1, 2017.
[12] F. Glover, "Tabu searchpart i," *ORSA Journal on computing*, vol. 1, no. 3, pp. 190–206, 1989.
[13] AT&T. (2017) AT&T Data Center Locations. [Online]. Available: https://www.business.att.com/enterprise/Service/cloud/colocation/data-center-locations
[14] WonderNetwork. (2017) Global Ping Statistics. [Online]. Available: https://wondernetwork.com/pings